\documentclass[usegraphicx,useAMS,usenatbib]{mn2e}

\usepackage{amssymb,amsmath} 
\usepackage{multirow}
\usepackage{sebs_journal_abrvs}

\usepackage{color}
\newcommand{\CHANGED}{}

\title[Circumbinary Orbits]{The Dynamics and Stability of Circumbinary Orbits}
  
\author[Doolin \& Blundell]{Samuel Doolin$^1$ and Katherine M.\ Blundell$^1$ \\
	$^1$University of Oxford, Department of Physics, Keble Road, Oxford, OX1 3RH, U.K.}

\begin{document}

\maketitle

\begin{abstract}

We numerically investigate the dynamics of orbits in 3D circumbinary phase-space as a function of binary eccentricity and mass fraction. We find that inclined circumbinary orbits in the elliptically-restricted three-body problem display a nodal libration mechanism in the longitude of the ascending node and in the inclination to the plane of the binary. We (i) analyse and quantify the behaviour of these orbits with reference to analytical work performed by \citet{Farago:2010} and (ii) investigate the stability of these orbits over time. This work is the first dynamically aware analysis of the stability of circumbinary orbits across both binary mass fraction and binary eccentricity. This work also has implications for exoplanetary astronomy in the existence and determination of stable orbits around binary systems.

\end{abstract}


\section{Introduction}

The recent discovery of a circumbinary disk around the microquasar SS433 \citep{Blundell:2008} along with evidence that such disks may be dynamically coupled to the accretion and outflow from such systems \citep{Artymowicz:1996,Regos:2005,Doolin:2009,Perez:2010} has led us to investigate the behaviour of orbits encompassing binary systems.

Additionally, in this era of exoplanetary astronomy we are finding new planets almost every week, but only a minority those so far discovered are in binary systems (e.g.~\citealp{Lee:2009,Beuermann:2011,Qian:2011}). With an unknown but significant fraction of stars confined to binaries, and with new technologies and methods to detect circumbinary planets (e.g.~\citealp{Schwarz:2011}), it is crucial that we understand the dynamics and stability of circumbinary orbits.

In the following sections we investigate a circumbinary nodal libration with the aid of finely time-sampled numerical studies. We compare our results to analytic work performed by \citet{Farago:2010} and investigate the accuracy and limits of their model.

With a second suite of numerical simulations we then determine the stability of the librating circumbinary orbits as a function of binary eccentricity and mass fraction.

\section{Methods and terminology}

\subsection{Orbital Elements}

\begin{figure}
	\centering
	\includegraphics[width=0.95\columnwidth]{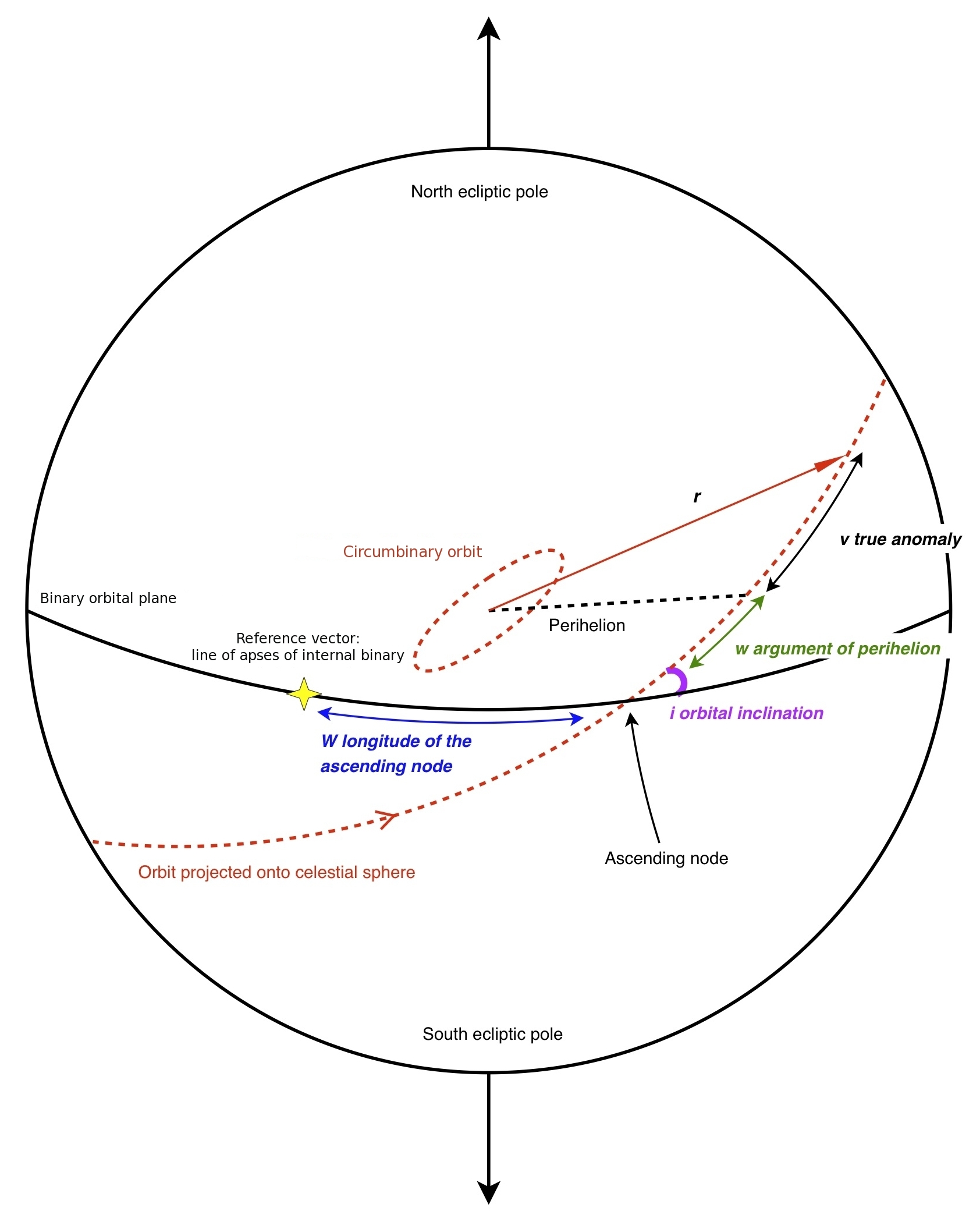}
	\caption{A representation of the Keplerian orbital elements.}
	\label{fig:3D}
\end{figure}

A general orbit in 3D is described intuitively by the Keplerian orbital elements \mbox{($a$, $e$, $i$, $W$, $w$, $v$)} illustrated by Figure~\ref{fig:3D}. The eccentricity $e$, semi-major axis $a$ and true anomaly $v$ describe the motion of a body in its orbital plane, whilst the inclination $i$, longitude of the ascending node $W$ and argument of perihelion $w$ describe the orientation of the orbital plane with respect to some reference plane and direction. The relationship between these orbital elements and a body's state vector (position and velocity) may be found in \citet{Green:1985} amongst others.

In this coordinate space all trajectories are represented uniquely, with a {\CHANGED Keplerian orbit} having the property of conserving all quantities but the true anomaly $v(t)$, which describes the exact position of a body on its orbital path.


In the circumbinary case we define the reference plane as the natural plane of the internal binary, and the reference direction as the vector along the binary's line of apses\footnote{i.e. the binary system's semi-major axis}. In our studies we consider the \emph{osculating} orbital elements of test particles --- that is to say their instantaneous orbital elements about the barycentre of the binary system --- as these quantities are no longer conserved.

\subsection{Numerical setup} \label{sec:numerial setup}

We have performed 3D numerical simulations of initially circular ($e=0$) circumbinary\footnote{`P-type' \citep{Dvorak:1989}} test particles in the elliptically restricted three-body problem. These massless test particles orbit synthetic binary systems of total mass $1M_{\odot}$ and semi-major axis $a_{\rm b}$ = 1AU. We scale all distances to the binary semi-major axis $a_{\rm b}$, and all times to the orbital period of the binary system $T_{\rm b}$.

We apply a customised adaptive step-size fourth and fifth-order Runge-Kutta integrator to integrate suites of test particles as a function of binary eccentricity $e_{\rm b}$ and mass fraction $\alpha_{\rm b}=m_1/(m_1+m_2)$ where $m_1\leq m_2$.

We track the orbital elements of each test particle about the centre of mass of the binary during integration, outputting time-lapsed snapshots to a database. We take advantage of the speed, organisation and SQL functionality of the database to study and accurately fit to the behaviours of orbits which we present in further sections. 

Each test particle is also monitored for instability. Unstable orbits are identified and removed during integration where a test particle is perturbed sufficiently from its initial orbit to approach either star, or if it evolves onto an unbound trajectory ($e>1$). {\CHANGED This paper is not concerned with the ultimate fate of any test particle that experiences a close encounter with either stellar body. These test particles are rejected from the simulations and the precise scattering is not computed. }

Post-simulation stability criteria are applied to identify and reject test particles which do not quite reach escape velocity.

\section{The nodal libration} \label{sec:dynamics}

\subsection{Suite of simulations}

Our first suite of simulations was designed to be extensively time-sampled to expose the dynamics of circumbinary orbits. The binary eccentricity and mass-fraction parameter space that we explore is laid out in Table~\ref{table:binary-space sampling} whilst the initial phase-space sampling of test particles is laid out in Table~\ref{table:dynamics sims phase-space sampling}. One additional spherical shell of test particles at a semi-major axis $50a_{\rm b}$ was also integrated for $5\times10^6T_{\rm b}$.

\begin{table}
\caption{Sampling of binary eccentricity and mass fraction}
\label{table:binary-space sampling}
\begin{center} 
\begin{tabular}{c c c c}
Orbital Element & min & max & $\triangle$ \\ 
\hline
eccentricity $e_{\rm b}$ & 0 & 0.6 & 0.1 \\
mass fraction $\alpha_{\rm b}$ & 0.1 & 0.5 & 0.1
\end{tabular}
\end{center} 
\end{table}

\begin{table}
\caption{Sampling of circumbinary phase space where \newline $a_{\rm b}$ = binary semi-major axis and $T_{\rm b}$ = binary orbital period. }
\label{table:dynamics sims phase-space sampling}
\begin{center} 
\begin{tabular}{c c c c}
Orbital Element & min & max & $\triangle$ \\ 
\hline
semi-major axis $a$	& 1.5$a_{\rm b}$ & 10$a_{\rm b}$ & 0.5$a_{\rm b}$\\
inclination $i$ & $0$ & $\pi$ & $\pi/20$ \\
longitude of the ascending node $W$ & $0$ & $2\pi$ & $\pi/2$\\
true anomaly $v$ & $0$ & $2\pi$ & $\pi/2$ \\
\hline
simulation length and snapshot $\triangle t$ &  & $10^4T_{\rm b}$ & $10T_{\rm b}$ 
\end{tabular}
\end{center} 
\end{table}

\subsection{Description} \label{sec:description}

As a typical test particle is integrated over the course of a simulation its semimajor axis $a$ and eccentricity $e$ remain constant. We discover a nodal libration mechanism in the inclination $i$ and longitude of the ascending node $W$ of all inclined orbits. 

\begin{figure*}
	\centering
	\includegraphics[height=0.9\textheight]{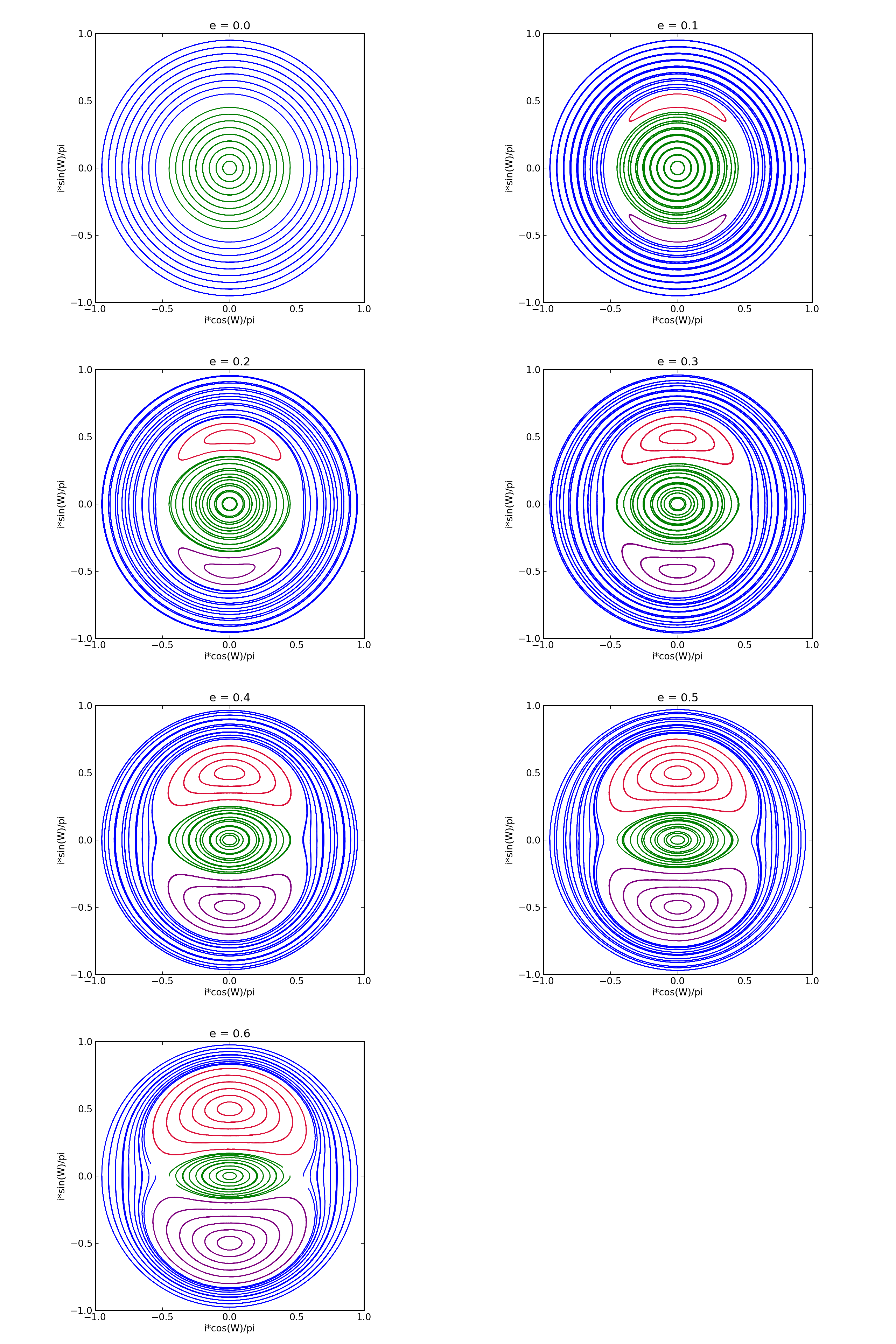}
	\caption{The $(i\cos{W},i\sin{W})$ surface of section for circumbinary orbits as a function of binary orbital eccentricity $e_{\rm b}$. \newline
	Green: prograde ($i<\pi/2$). Precession is clockwise. \newline 
	Blue: retrograde ($i>\pi/2$). Precession is anti-clockwise. \newline
	Red: island of libration centred at ($i=\pi/2,W=\pi/2$). Precession is anti-clockwise. \newline
	Purple: Island of libration centred at ($i=\pi/2,W=-\pi/2$). Precession is anti-clockwise.}
	\label{fig:surface of section}
\end{figure*}

This is best visualised in a polar $(i\cos{W},i\sin{W})$ slice through parameter space, which may be termed a surface of section, as illustrated by Figure~\ref{fig:surface of section}. Each line in Figure~\ref{fig:surface of section} is traced out by an individual test particle over the course of a simulation. Figure~\ref{fig:surface of section} reveals four distinct populations that we now examine in turn.

\subsubsection{Prograde orbits \emph{(green)}}

An inclination $i=0$ corresponds to a coplanar prograde circumbinary orbit. Orbits of inclination $0\leq i<\pi/2$, whilst not necessarily coplanar, we refer to as prograde. The prograde region of phase space therefore extends out from the centre ($i=0$) of the surface of section (Figure~\ref{fig:surface of section}) and encompasses all orbits of a similar behaviour.

Prograde orbits exhibit a precession in the longitude of the ascending node $W$. This evolution in $W$ produces clockwise paths around the surface of section shown in Figure~\ref{fig:surface of section}. 


\subsubsection{Retrograde orbits \emph{(blue)}}

An inclination $i=\pi$ corresponds to a coplanar retrograde circumbinary orbit. Orbits of inclination $\pi/2 < i \leq \pi$, whilst not necessarily coplanar, we refer to as retrograde. The retrograde region of phase space therefore extends inwards from the outer limit ($i=\pi$) of the surface of section (Figure~\ref{fig:surface of section}) and encompasses all orbits of a similar behaviour.

Retrograde orbits exhibit a precession in the longitude of the ascending node $W$. But counter to the prograde orbits the retrograde evolution in $W$ produces anti-clockwise paths around the surface of section (Figure~\ref{fig:surface of section}). 


{\CHANGED We expect that the precession in $W$ observed in close-to-coplanar prograde and retrograde orbits is due to a coupling between the specific angular momentum of test particles on inclined orbits and the $\hat{z}$ angular momentum of the binary. Such a coupling would exert a torque on the test particle producing a precession in the ascending node, akin to gyroscopic precession.}

\subsubsection{Islands of libration  \emph{(red \& purple)}}

An inclination $i=\pi/2$ corresponds to a circumbinary orbit which is exactly perpendicular to the binary plane. Figure~\ref{fig:surface of section} shows two very clear libration islands centred on \hbox{$i=\pi/2$, $W=\pm\pi/2$}. A test particle on an orbit within a region of libration has its inclination $i$ and ascending node $W$ coupled to precess about the centre of libration. For both regions of libration this precession is anti-clockwise.

\subsection{The geometry of the surface of section} \label{sec:geometry}

The geometry of the $(i\cos{W},i\sin{W})$ surfaces of section shown in Figure~\ref{fig:surface of section} reveal a dependence on the internal binary eccentricity. More specifically, the extent of the two regions of libration can be seen to scale with binary eccentricity. A circular binary $e_{\rm b}=0$ exhibits no libration islands, whereas for a binary of $e_{\rm b}=0.6$ the libration mechanism is becoming the dominant behaviour in phase space. We quantify the extent of the libration regions in \S~\ref{sec:separatrix and critical angle}.

Inspection of surfaces of section across values of binary mass fraction  $\alpha_{\rm b}$ and radius $a$ lead us to conclude that the geometry is both mass fraction and radius independent. The period of the precession of each test particle however does show a strong dependance on $\alpha_{\rm b}$ and $a$, which we explore in \S~\ref{sec:period}.

\subsubsection{Kozai cycles}

The geometry of the polar ($i$,$W$) surface of section (Figure~\ref{fig:surface of section}) with its islands of libration appears similar to that of the Kozai mechanism.  \citet{Kozai:1962} showed analytically that an inclined \emph{circumstellar}\footnote{An orbit about one star of a binary system. `S-type' \citep{Dvorak:1989}.} orbit may experience an oscillating exchange between inclination and eccentricity, and also a libration in the argument of perihelion. But whilst the \emph{circumstellar} ($i$,$w$) plane may share similarities with our \emph{circumbinary} ($i$,$W$) plane, we are dealing with a very different regime of the elliptically-restricted three-body problem.

\subsubsection{Symmetry} \label{sec:symmetry}

The $(i\cos{W},i\sin{W})$ surface of section (Figure~\ref{fig:surface of section}) is a 2D projection of the surface of a unit sphere, where $i$ corresponds to the polar angle and $W$ to the azimuthal angle. The central point of the surface of section ($i=0$) corresponds to a co-planar prograde orbit (see Figure~\ref{fig:3D}) and at this point the longitude of the ascending node $W$ is undefined. Equivalently one may regard this $i=0$ point as situated at the north pole of a sphere, with the south pole at $i=\pi$ (a coplanar, retrograde orbit). Points of $i=\pi/2$ identify the equator. This type of spherical projection is known in cartographic circles as the azimuthal equidistant projection.

\begin{figure}
	\centering
	\includegraphics[width=\columnwidth]{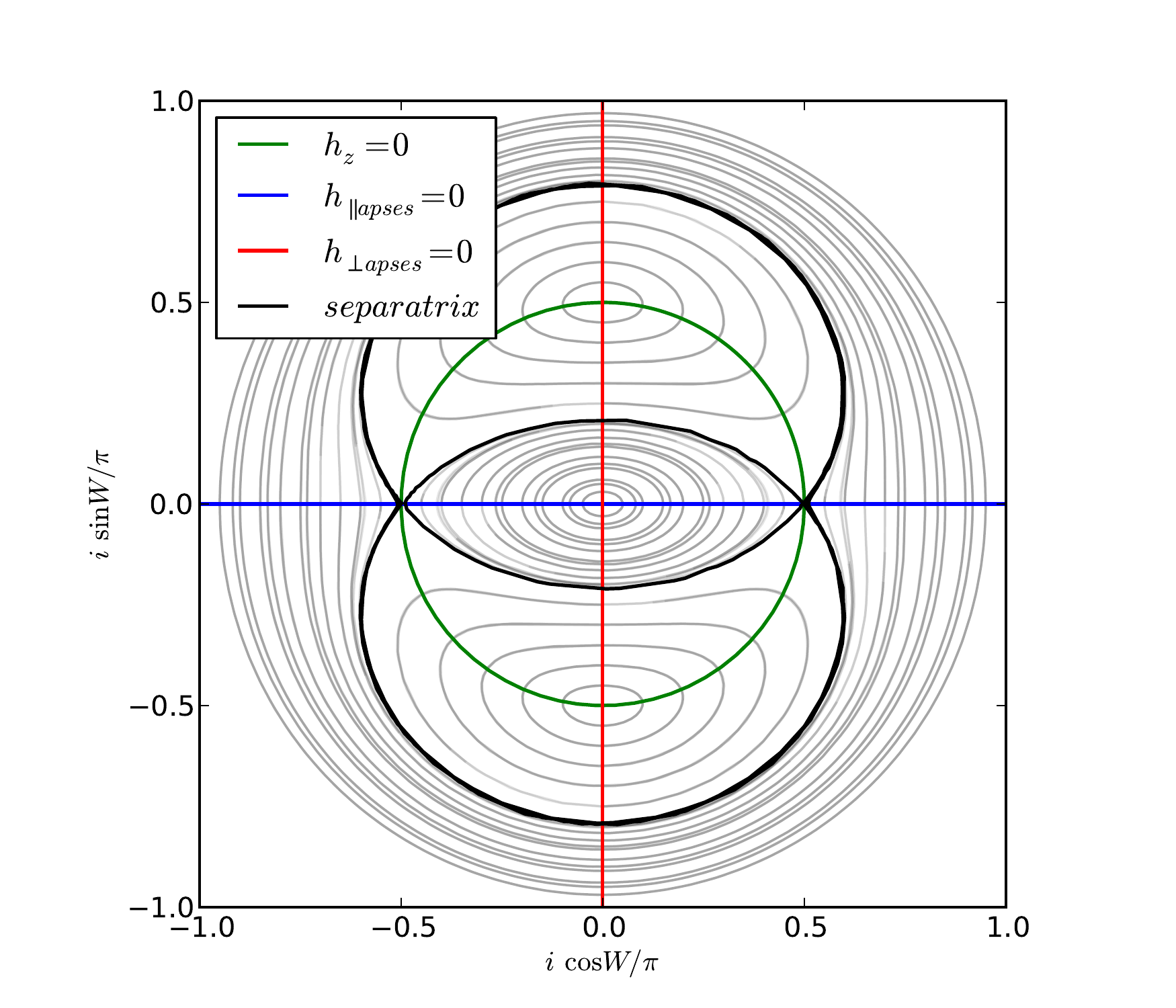}
	\caption{The separatrix and symmetries of the $(i\cos{W},i\sin{W})$ surface of section. $h =$ specific angular momentum. }
	\label{fig:hxhyhz}
\end{figure}

This unit sphere essentially defines the direction of the specific angular momentum vector $\vec{h}$ of a test particle. We note three particular planes of symmetry through this sphere. These may be specified by considering components of a circumbinary orbit's specific angular momentum $\vec{h}$ along axes (i) parallel to the line of apses of the binary, (ii) perpendicular to the line of apses of the binary\footnote{yet remaining in the binary orbital plane}, and (iii) perpendicular to the plane of the binary. We show these  three planes of symmetry  in Figure~\ref{fig:hxhyhz} via the colours:
\\ ~ \\
\begin{tabular}{ l l l }
\bf green:	& $i=\pi/2$				& $h_z=0$ \\
\bf blue:	& $W \in \{ 0 , \pi \}$			& $h_{\parallel {\rm apses}}=0$ \\
\bf red:	& $W \in \{ \pi /2 , 3 \pi / 2 \}$	& $h_{\perp {\rm apses}}=0$
\end{tabular}

\subsection{Previous work}

Confirmation that the features of Figure~\ref{fig:surface of section} are not numerical artefacts comes reports of from similar behaviour in \citet{Verrier:2009}. In that work the authors report discovering a counter-play between the Kozai mechanism and a new circumbinary libration whilst modelling orbits within the double binary system HD98800.

Following on from \citet{Verrier:2009} is the excellent analytic paper of \citet{Farago:2010}. In this article the authors consider the elliptically restricted three-body problem and take advantage of the assumption that, in the circumbinary case, the displacement of the third body $r_{3}$ is greater than the relative separation of the binary $r_{21}$.

Under this $r_{3} \gg r_{21}$ approximation \citeauthor{Farago:2010} expand the three body Hamiltonian to second order in $r_{21}/r_3$ and then average over the orbit of the binary and the third body to obtain a time-averaged quadrupolar Hamiltonian.

This Hamiltonian promises to be very accurate in the regime $r_{21}/r_{3} \ll 1$, as higher order terms in $r_{21}/r_3$ will tend to zero at a faster rate than those of second order. This model should not be so good for orbits closer to the binary system, where higher order terms will play a more substantial role. In the following sections we investigate the accuracy of \citeauthor{Farago:2010}'s model by testing the predictions that it makes, and the limits at which their quadrupolar approximation becomes insufficient.

\subsection{Separatrix and critical angle} \label{sec:separatrix and critical angle}

The separatrix (Figure~\ref{fig:hxhyhz}: black) is the boundary in phase-space between different modes of behaviour. In our circumbinary surface of section the separatrix takes the form of a triple figure-of-eight, or two circles intersecting at ($W=0$, $i=\pi/2$) and ($W=\pi$, $i=\pi/2$), dividing the regions of behaviour outlined in \S~\ref{sec:description} above. The separatrix which we plot in Figure~\ref{fig:hxhyhz} is actually the path of a rare test particle which, due to our stepping integrator, sampled more than one region of behaviour.

In Section \ref{sec:geometry} we mentioned that the geometry of the surface of section is predominantly dependent on binary eccentricity, and here we quantify this. We note that each point on the surface of section (Figure~\ref{fig:surface of section}) defines a unique path, and that every path intersects the vertical axis ($W=\pm\pi/2$). We define a critical angle $i_{\rm crit}$ as the inclination $i$ at which the separatrix crosses the positive vertical axis ($W=+\pi/2$) in the region $0\leq i \leq \pi/2$. 

Subsequent to the discussion of symmetry in \S~\ref{sec:symmetry} it follows that the three other intersections of the separatrix with the vertical axis are reflections of the critical angle $i_{\rm crit}$ defined in the region \hbox{($W=\pi/2$, $0\leq i \leq \pi/2$)}.

\subsubsection{Measuring the critical angle}

We have run an additional suite of simulations to find and extract the critical angle as a function of binary eccentricity. These simulations were run with one shell of test particles at radius $25a_{\rm b}$, longitude of the ascending node $W=\pi/2$ and with a high resolution in inclination, intervals of $\triangle i = \pi / 80$, at various values of binary eccentricity. 

\begin{figure}
	\centering
	\includegraphics[width=\columnwidth]{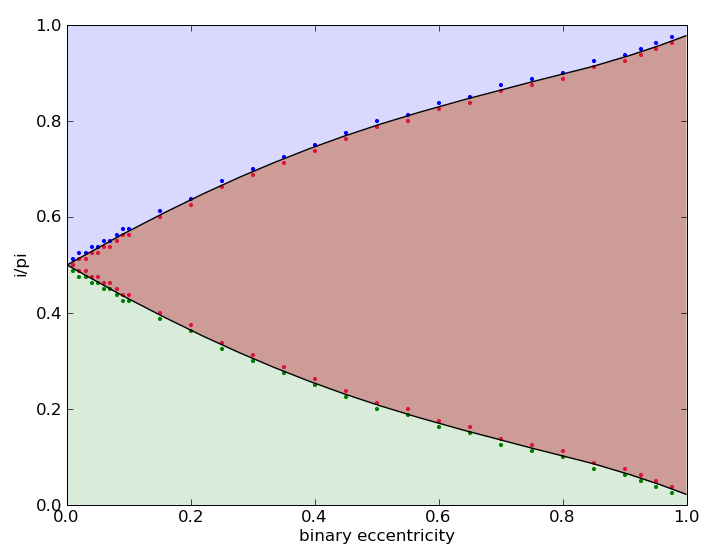}
	\caption{Orbital behaviour in the $W=\pi/2$ plane as a function of binary eccentricity. \newline
	Green: prograde ($i<\pi/2$) \newline 
	Blue: retrograde ($i>\pi/2$) \newline
	Red: island of libration centred at ($i=\pi/2,W=\pi/2$) 
	}
	\label{fig:L1}
\end{figure}

For each orbit sampled we visually identify the behaviour (\S~\ref{sec:dynamics}) as a function of inclination and binary eccentricity to find upper and lower boundaries on the separatrix. The results are plotted in Figure~\ref{fig:L1} preserving the colour scheme of Figure \ref{fig:surface of section}.

The green, red and blue points in Figure~\ref{fig:L1} represent the behaviour of test particles at these locations. We only plot the sampled points that lie either side of the separatrix. We extract the simplest polynomial fit to accurately describe the critical angle ($0\leq i_{\rm crit} \leq \pi/2$) as a function of binary eccentricity $e_{\rm b}$ with the constraint that the fit must pass between every pair of green-red points. This is given by:\

\begin{equation} \label{eq:i_crit}
i_{\rm crit} = 0.5 + ae_{\rm b} + be_{\rm b}^2 + ce_{\rm b}^3 + de_{\rm b}^4,
\end{equation}

where
\begin{align*}
&a = -0.7138 \pm 0.0023 \\
&b = 0.1021  \pm 0.0030 \\
&c = 0.5264  \pm 0.0038 \\
&d = -0.3942 \pm 0.0047.
\end{align*}

In all cases the angle of the separatrix in the region ($\pi/2 \leq i_{\rm crit} \leq \pi$) is at inclination $\pi- i_{\rm crit}$, as it should be by symmetry arguments. 

\subsubsection{Comparison with \citet{Farago:2010}}

The time-averaged quadrupolar model of \citet{Farago:2010} predicts the critical angle (their equation 2.34) and hence the location of the separatrix as

\begin{equation} \label{eq:i_crit_farago}
i_{\rm crit} = \arcsin{\sqrt{\frac{1-e_{\rm b}^2}{1+4e_{\rm b}^2}}}
\end{equation}

Since we measure the critical angle at a radius of $25a_{\rm b}$ from the binary we expect the quadrupolar approximation to hold, and indeed we find an excellent agreement between our measurements and Equation~\ref{eq:i_crit_farago}. The predicted location of the separatrix lies between every pair of our experimental limiting points.

\begin{figure}
	\centering
	\includegraphics[width=\columnwidth]{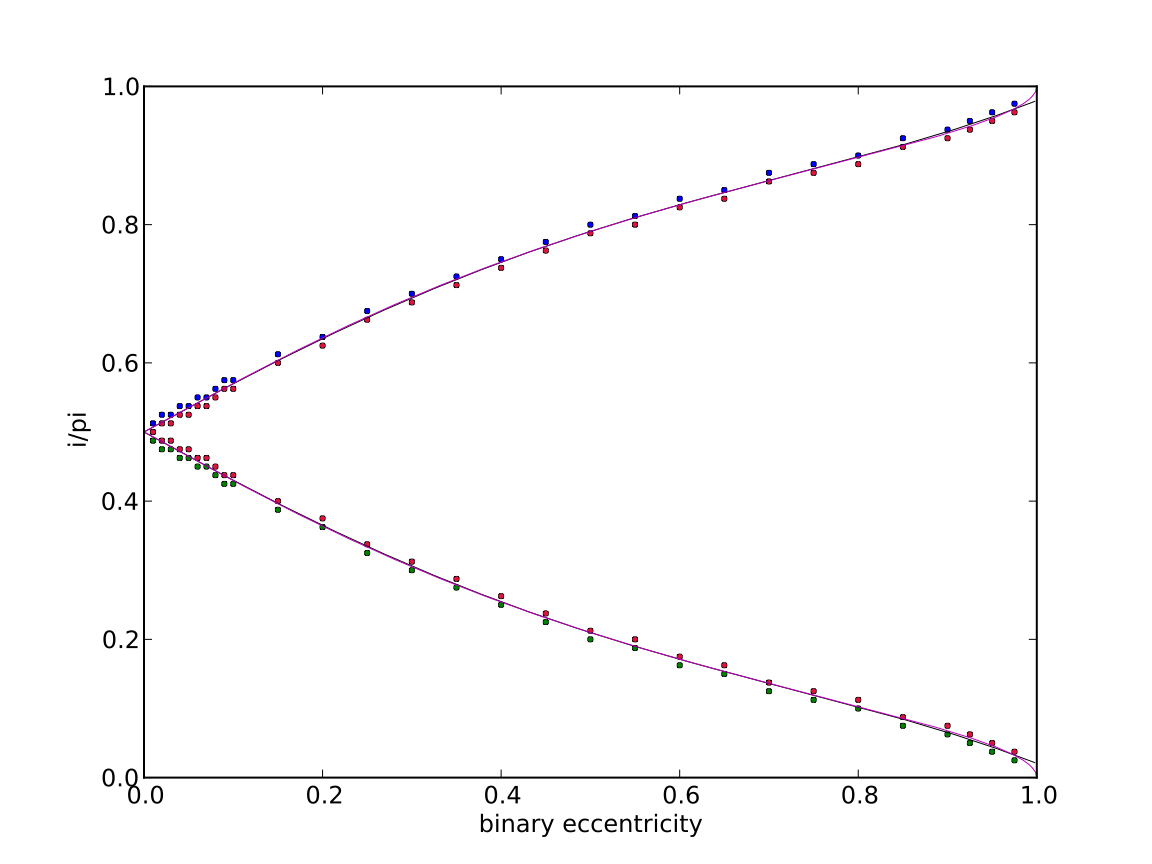}
	\caption{Our experimental fit to the critical angle of the separatrix (Eq~\ref{eq:i_crit}) (black) and \citeauthor{Farago:2010}'s analytic expression (Eq~\ref{eq:i_crit_farago}) (pink).  \newline
	Green: prograde ($i<\pi/2$) \newline 
	Blue: retrograde ($i>\pi/2$) \newline
	Red: island of libration centred at ($i=\pi/2,W=\pi/2$) 
	}
	\label{fig:separatrix_farago}
\end{figure}

We plot our data, fit and the \citeauthor{Farago:2010} prediction together in Figure~\ref{fig:separatrix_farago}. There is an almost exact agreement between model and data at this radius. The largest divergence is at $e_{\rm b} \rightarrow 1$, which is an unphysical limit as the binary itself becomes unbound.

\subsection{Constant of motion}

\citet{Verrier:2009} proposed an integral of motion (see their equation 3) for the libration islands to be the component of the specific angular momentum of an orbit along the line of apses of the internal binary, which may be expressed as
\begin{equation} \label{eq:verrier_h}
h_{\parallel {\rm apses}} = h\sin{i}\sin{W}.
\end{equation}

This makes intuitive sense as the libration islands are centred at \hbox{($i=\pi/2$, $W=\pm\pi/2$)}, i.e. the points at which a test particle's angular momentum is exactly parallel or antiparallel to the line of apses of the internal binary (see Figure~\ref{fig:hxhyhz}). So for small deviations from these central points we find that $h_{\parallel {\rm apses}}$ is conserved.

\begin{figure}
	\centering
	\includegraphics[width=\columnwidth]{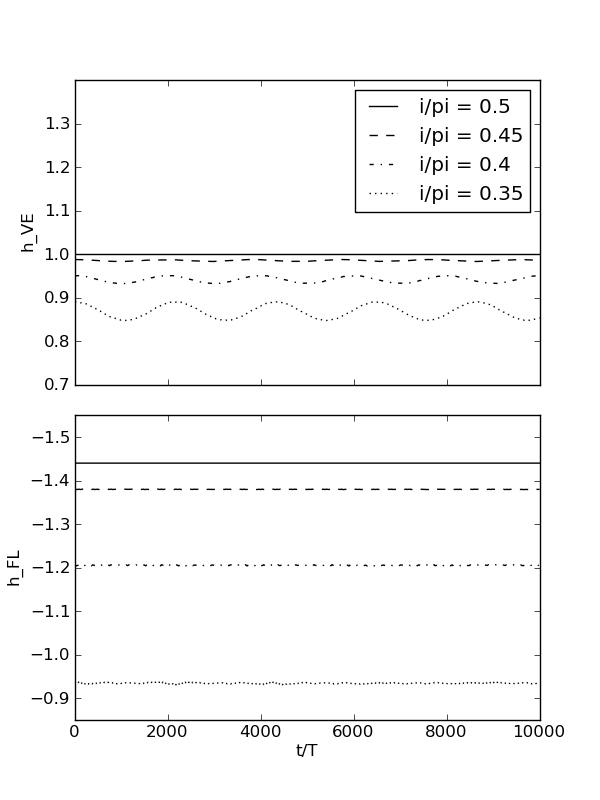}
	\caption{A plot of the suggested constants of motion for example test particles in proximity to the centre of the island of libration ($i=\pi/2$, $W=\pi/2$) from our simulation of binary eccentricity $e_{\rm b} =$ 0.6 and mass fraction $\alpha_{\rm b} =$ 0.5. Upper panel: \citet{Verrier:2009} (Eq~\ref{eq:verrier_h}), lower panel:  \citet{Farago:2010} (Eq~\ref{eq:h_farago}).}
	\label{fig:CM_t}
\end{figure}

Unfortunately this model breaks down as we move out from the centre of libration. In Figure~\ref{fig:CM_t} (upper panel) we show the \citeauthor{Farago:2010} constant of motion (Eq~\ref{eq:verrier_h}) over the course of our integration for example test particles in proximity to the centre of the island of libration ($i=\pi/2$, $W=\pi/2$) from our simulation of binary eccentricity $e_{\rm b} =$ 0.6 and mass fraction $\alpha_{\rm b} =$ 0.5. We observe that this suggested constant of motion becomes insufficient for test particles of lower inclination, which sample phase space away from the centre of the libration.

A more promising integral of motion is given by the time-averaged quadrupolar model of \citet{Farago:2010}, their equation 2.20, which we translate to orbital elements as
\begin{equation} \label{eq:h_farago}
h_{\rm FL} = \cos^2{i} - e_{\rm b}^2\sin^2{i}(5\sin^2{W}-1).
\end{equation}

We see that this equation reduces to the square of Equation~\ref{eq:verrier_h} for values of ($i\rightarrow\pi/2$, $W\rightarrow\pm\pi/2$) close to the centre of libration. But this model correctly predicts the paths of orbits across the entirety of the surface of section and for values of binary eccentricity $e_{\rm b}$. We plot this constant of motion in the lower panel of Figure~\ref{fig:CM_t} for the same test particles as above. 

We use the constant of motion predicted by the time-averaged quadrupolar model of \citet{Farago:2010} as a test of the limits of the quadrupolar approximation. For each simulated test particle we have $10^3$ snapshots over the course of integration. We therefore calculate an instantaneous $h_{\rm FL}$ for each test particle at each snapshot and ask the question: {\it `how constant is the constant of motion?'}

For example --- in our shell of test particles at radius $50a_{\rm b}$ we find that a typical test particle experiences variation in $h_{\rm FL}$ of $\sim 1.4\times 10^{-5}$. This is measured by taking the standard deviation $\sigma$ of the instantaneous measurements of $h_{\rm FL}$ over the course of a simulation. Since $h_{\rm FL}$ is of order unity, this equates to a typical error of 0.0014\% at these large radii.

\begin{figure}
	\centering
	\includegraphics[width=\columnwidth]{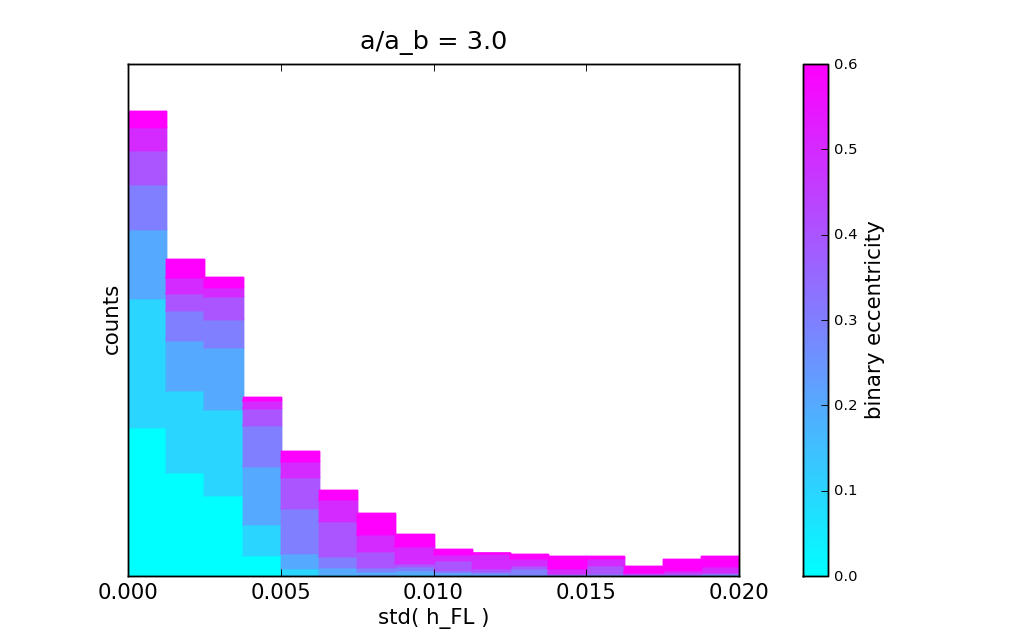}
	\includegraphics[width=\columnwidth]{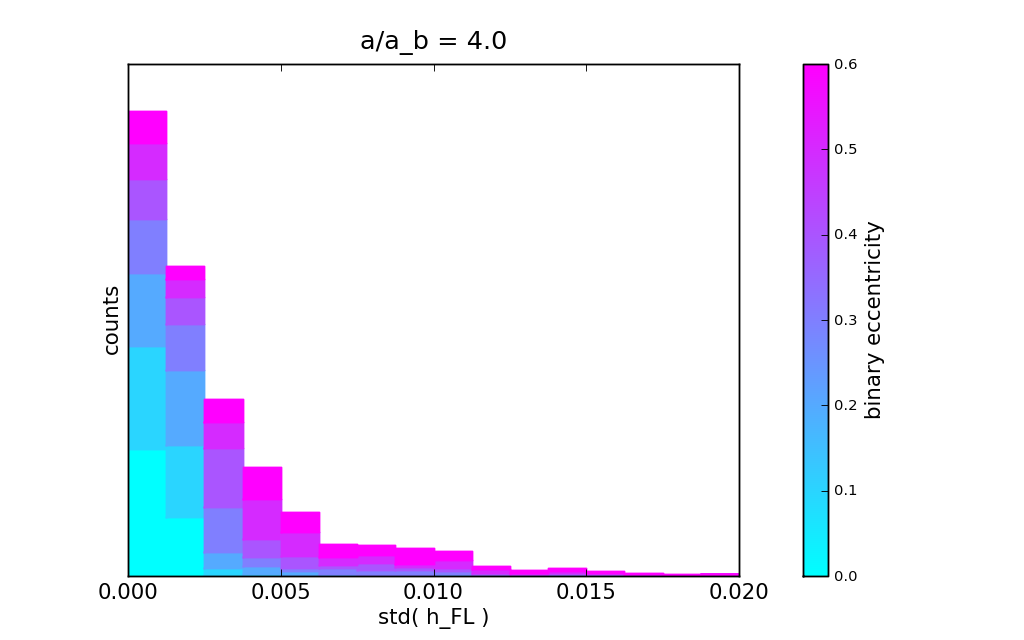}
	\includegraphics[width=\columnwidth]{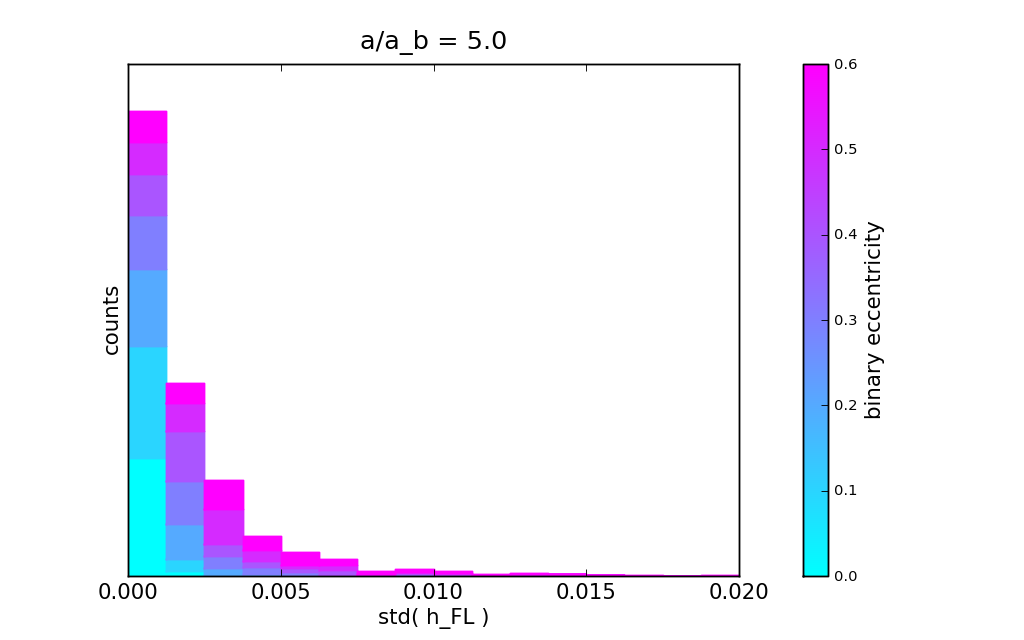}
	\includegraphics[width=\columnwidth]{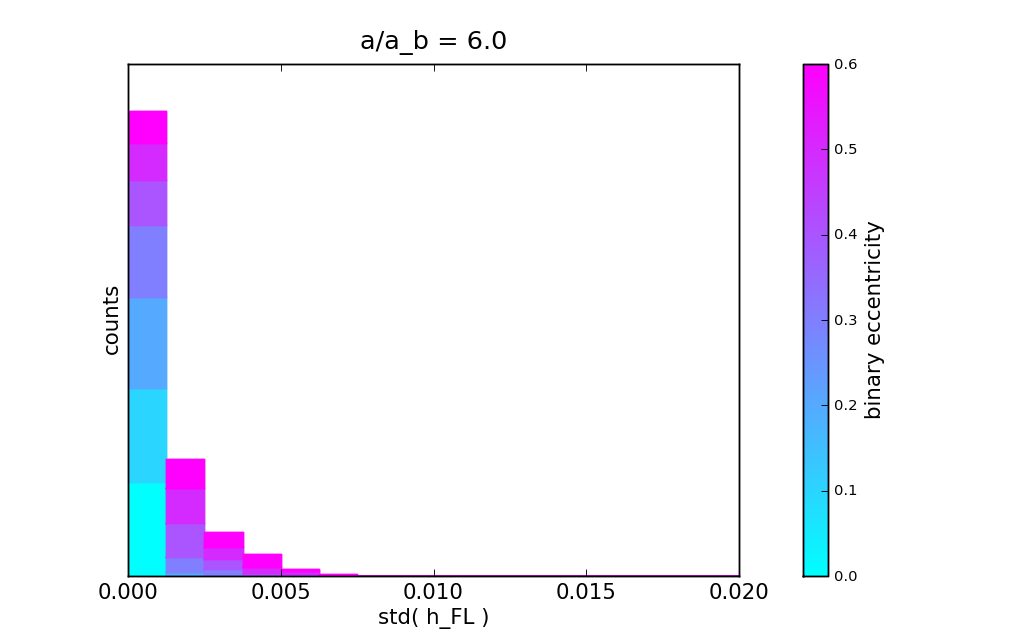}
	\caption{Histograms of $\sigma(h_{\rm FL})$ for radii 3, 4, 5 and 6$a_{\rm b}$, coloured according to binary eccentricity $e_{\rm b}$.}
	\label{fig:h_farago}
\end{figure}

We now investigate how well the quadrupolar approximation holds as we consider orbits closer to the binary. To do so we create histograms of $\sigma(h_{\rm FL})$ for each shell of test particles that we sample (see Table~\ref{table:dynamics sims phase-space sampling}). In Figure~\ref{fig:h_farago} we show histograms of $\sigma(h_{\rm FL})$ for radii 3, 4, 5 and 6$a_{\rm b}$. We also colour the histograms according to the simulated binary eccentricity $e_{\rm b}$.

As we consider orbits closer to the binary ($a<10a_{\rm b}$), we see that the quadrupolar approximation begins to break down. At 6$a_{\rm b}$ most orbits are accurate to the quadrupolar model to better than 0.5\%, but as we move in to 3$a_{\rm b}$ some orbits experience deviation of $>1\%$. We observe similar small perturbations in a test particle's semi-major axis $a$ and its initially zero eccentricity $e$.

We also note that orbits around binary systems of higher eccentricity (Figure~\ref{fig:h_farago}: purple) deviate significantly more than those around circular binaries (Figure~\ref{fig:h_farago}: light blue), which indicates that the higher order terms of the \citeauthor{Farago:2010} Hamiltonian have a greater dependence on binary eccentricity.

{ \CHANGED

\subsubsection{Significance of $\sigma(h_{\rm FL})$} \label{sec:noise?}

In Figure~\ref{fig:noise?} we plot four typical examples of the variation in $h_{\rm FL}$ over the course of integration. These are selected from a random sample of 1000 such plots of test particles from the 3$a_{\rm b}$ radius bin (Figure~\ref{fig:h_farago}, upper panel).

Numerical noise from a stepping integrator would in principle accumulate over time. In Figure~\ref{fig:noise?} we show that the variation in $h_{\rm FL}$ is present from the start and constant in magnitude over the duration of a simulation. Values of $\sigma(h_{\rm FL})$ are therefore not biased by an accumulation of numerical noise towards the end of the simulation.

This absence of bias is additionally supported by the appearance of structure in Figure~\ref{fig:noise?},  most prominently in the lowest panel, which is further evidence of the physical effect of higher order terms than those present in the quadrupolar Hamiltonian of \citet{Farago:2010}.

\begin{figure}
	\centering
	\includegraphics[width=\columnwidth]{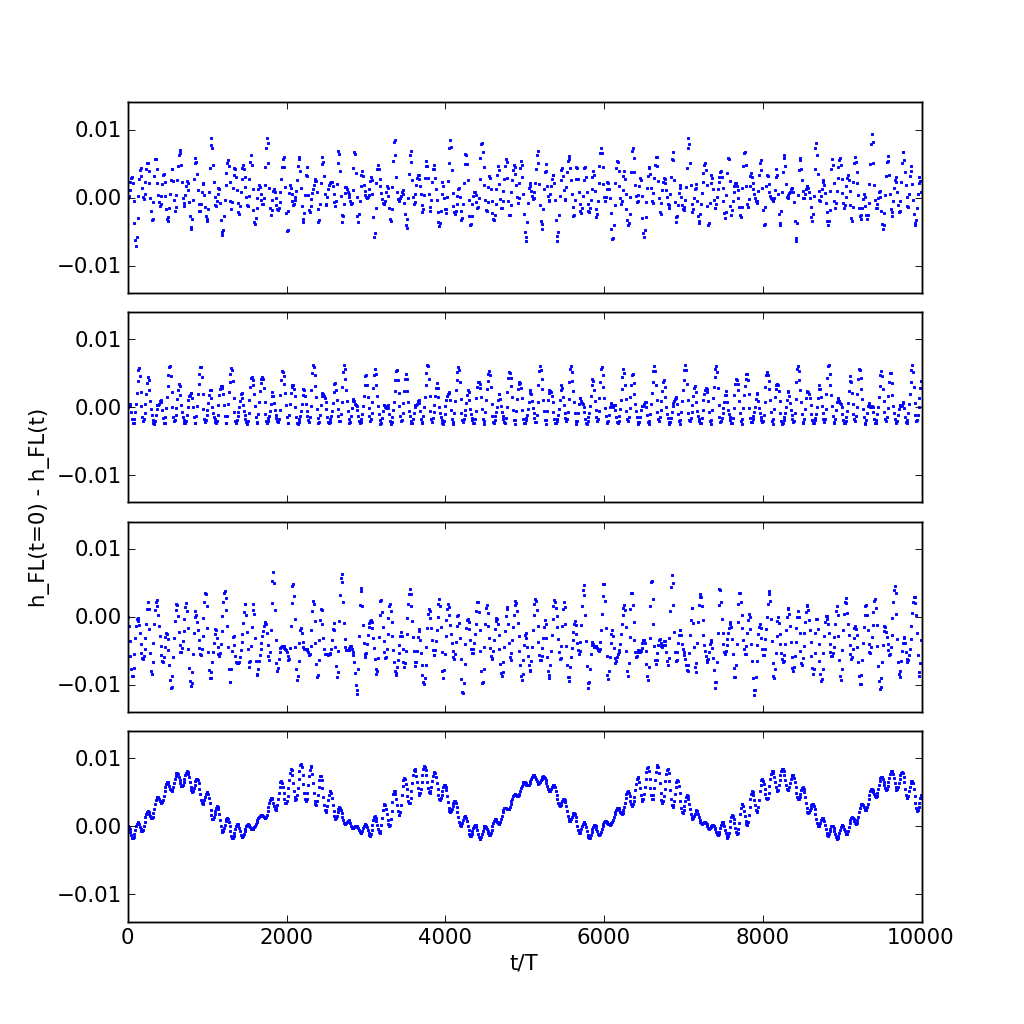}
	\caption{A plot of the variation in $h_{\rm FL}$ as a function of time for example test particles from the radius 3$a_{\rm b}$ bin.}
	\label{fig:noise?}
\end{figure}

} 

\subsection{Period of precession} \label{sec:period}

From our finely time-sampled $\triangle t = 10T_{\rm b}$ data (Table~\ref{table:dynamics sims phase-space sampling}) we extract a period $P$ for the precession of each stable test particle. These range from as little as $50T_{\rm b}$ (or $5\times\triangle t$) to greater than the simulation duration, but good fits are obtained for the vast majority.

\begin{figure*}
	\centering
	\includegraphics[height=0.9\textheight]{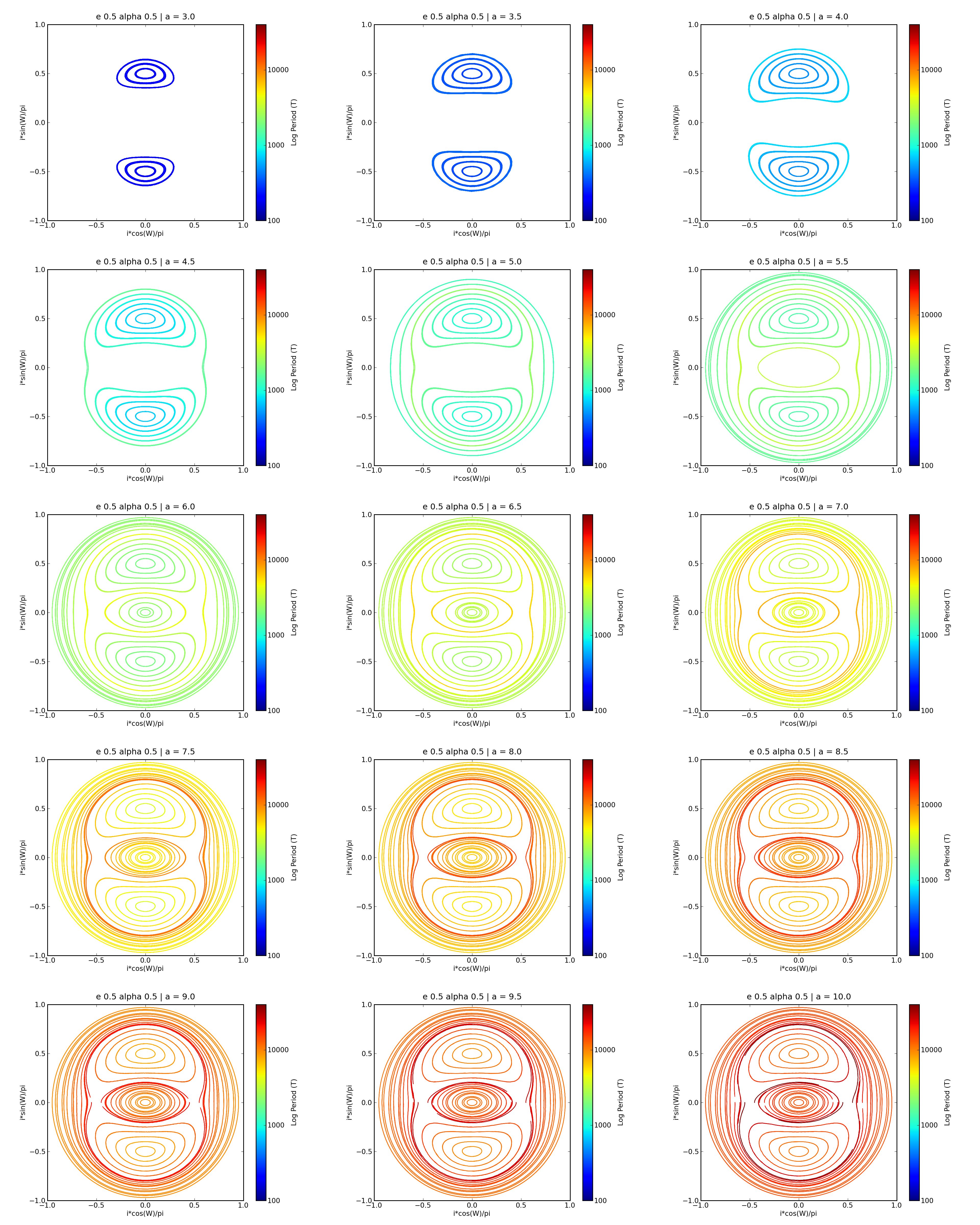}
	\caption{The $(i\cos{W},i\sin{W})$ surface of section for circumbinary orbits about a binary system of eccentricity $e_{\rm b}=0.5$ and mass fraction $\alpha_{\rm b} = 0.5$ at varius radii (scaled by binary semi-major axis $a_{\rm b}$), each track coloured by log period in binary orbital periods $T_{\rm b}$.}
	\label{fig:log_period}
\end{figure*}

In Figure~\ref{fig:log_period} we plot test particle traces on the $(i\cos{W},i\sin{W})$ surface of section for our simulation of binary eccentricity $e_{\rm b}=0.5$ and binary mass fraction $\alpha_{\rm b} = 0.5$, as a function of test particle radius $a/a_{\rm b}$, colouring the traces by precession period.

We observe that the period of precession $P$ is correlated with a test particle's orbital radius and proximity to a separatrix. Orbits which appear to be missing from Figure~\ref{fig:log_period} are unstable, and hence are not plotted. We explore the topic of stability in much greater detail in the second part of the paper (\S~\ref{sec:stability}).

\subsubsection{$P\propto a^n$}

\begin{figure}
	\centering
	\includegraphics[width=\columnwidth]{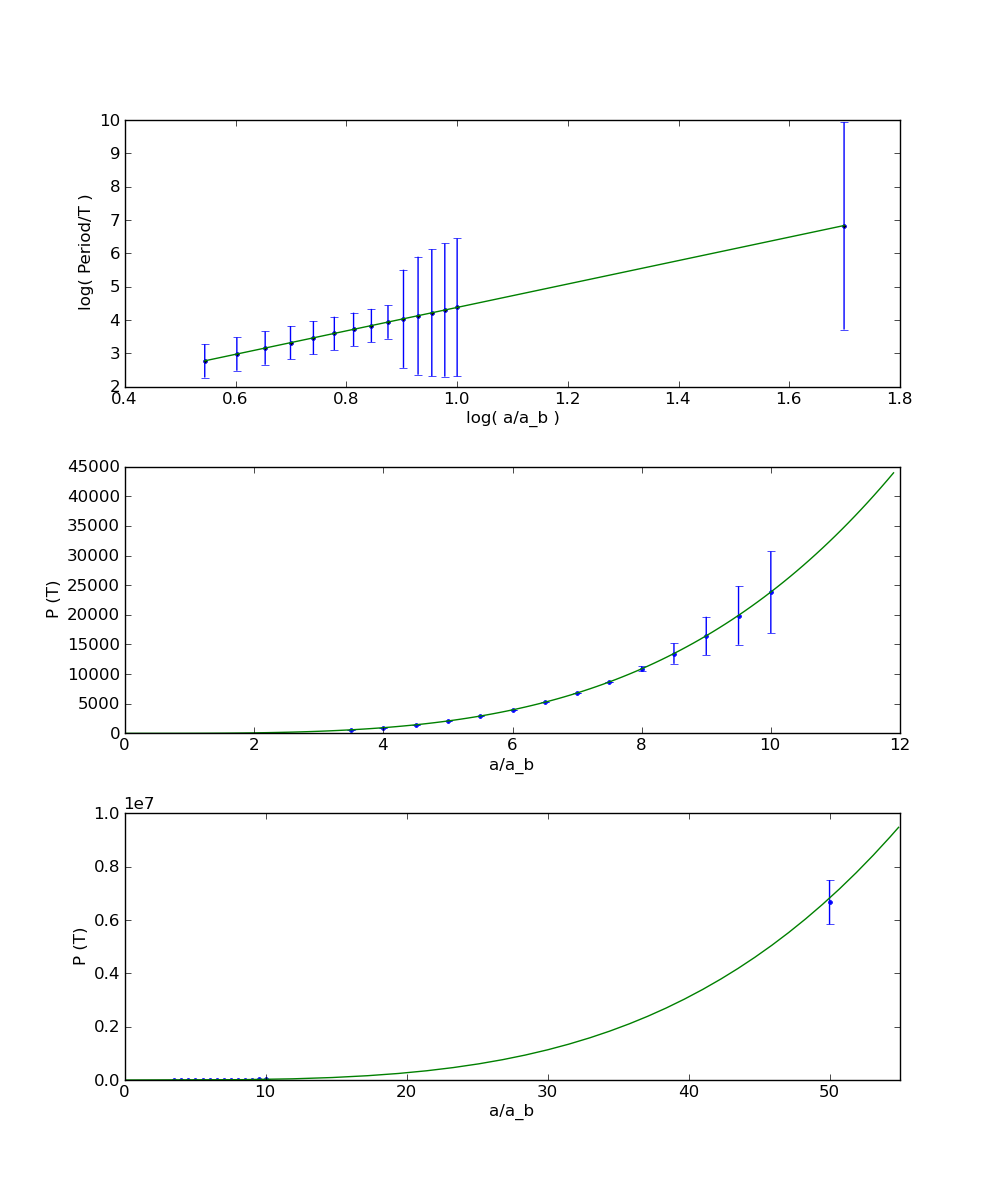}
	\caption{An example fit of $P\propto a^n$. This is for a prograde orbit of inclination $i=\pi/4$ about a binary of eccentricity $e_{\rm b}=0$ and mass fraction $\alpha_{\rm b}=0.5$. Plotted is precession period P (in units of binary orbital period $T$) vs test particle semi-major axis $a$ (in units of binary semi-major axis $a_{\rm b}$).}
	\label{fig:example fit}
\end{figure}

As discussed above and demonstrated by Figure~\ref{fig:log_period} the closer a circumbinary test particle orbits to the binary system the shorter is its period of precession $P$. We find that a powerlaw provides a very good fit to $P$ as a function of distance from the binary, and so we fit $P\propto a^n$ to all orbits sampled, where $n$ is a free parameter. An example fit is shown in Figure~\ref{fig:example fit}. The data point from our simulation at radius $50a_{\rm b}$ provides a very good constraint.

\begin{figure}
	\centering
	\includegraphics[width=\columnwidth]{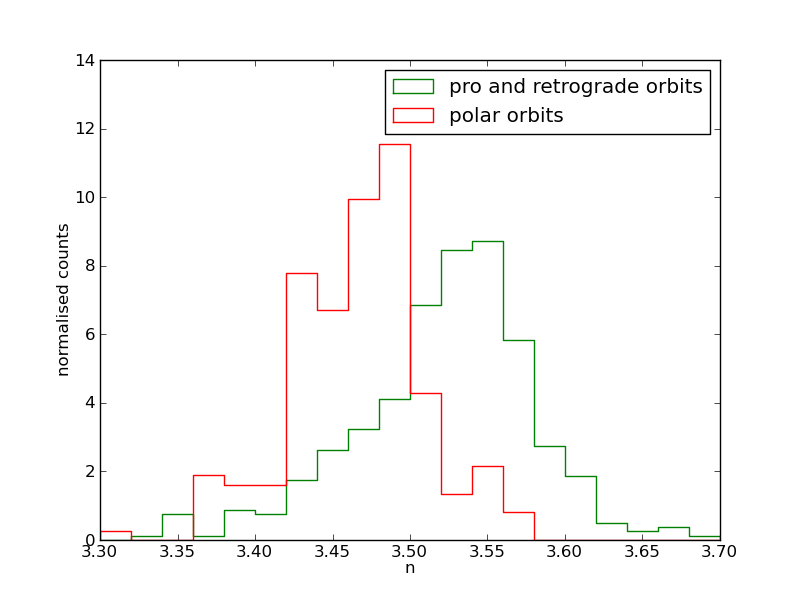}
	\caption{A histogram of fitted $n$ in $P\propto a^n$ showing two distinct populations: the polar orbits and the pro/retrograde orbits.}
	\label{fig:n_histogram}
\end{figure}

Our data show a tight clustering about $n=3.5$, as plotted in Figure~\ref{fig:n_histogram}. This is not within the range given by \citet{Verrier:2009} of $n=3.37\pm0.06$, but does agree with the time-averaged quadrupolar model of \citeauthor{Farago:2010} which predicts $n$ of exactly 3.5.

But on closer inspection of Figure~\ref{fig:n_histogram} we notice that $P\propto a^n$ appears to consist of two populations. One of these populations is comprised of the prograde and retrograde orbits, which cluster around $n=3.53$, and the other corresponds to the polar orbits, which cluster around $n=3.47$.

\subsubsection{An analytic expression for period of precession}

The time-averaged quadrupolar model of \citet{Farago:2010} makes the following prediction (their equation 2.32):

\begin{equation} \label{eq:T_farago}
\frac{P}{T_{\rm b}} = \frac{8}{3\pi}\frac{1}{\alpha_{\rm b}(1-\alpha_{\rm b})}{\Big (}\frac{a}{a_{\rm b}}{\Big )}^{7/2}\frac{F(k^2)(1-e^2)^2}{\sqrt{(1-e_{\rm b}^2)(h+4e_{\rm b}^2)}}
\end{equation}

\noindent where $h=h_{\rm FL}$ as defined in Eq~\ref{eq:h_farago},

$$ k^2 = \frac{5e_{\rm b}^2}{1-e_{\rm b}^2}\frac{1-h}{h+4e_{\rm b}^2} $$

\noindent and $F(k^2)$ may be defined in terms of the complete elliptical integral of the first kind, $K(k^2)$, as 

\begin{align*}
F(k^2) = {\Big \{ } { \> {K(k^2)} \atop \> {k^{-1} K(k^{-2})}  } & { {\rm where} \> k^2<1 \atop {\rm where} \> k^2>1}
\end{align*}

\noindent where
$$ K(k^2) = \int_{0}^{\pi/2} \frac{d\theta}{\sqrt{1-k^2\sin^2{\theta}}} .$$

\begin{figure}
	\centering
	\includegraphics[width=\columnwidth]{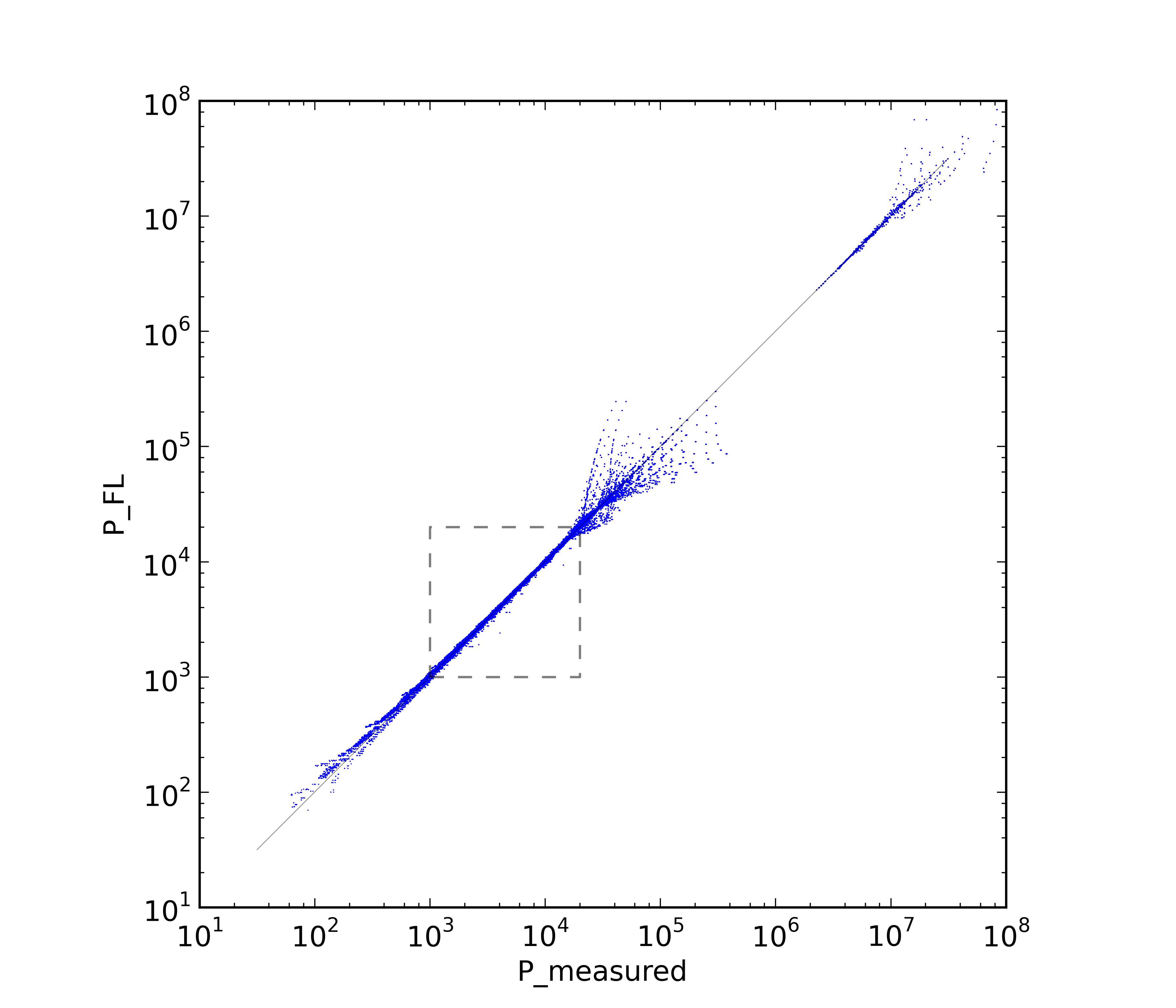}
	\caption{A plot of precession period $P$ in units of binary orbital period  $T_{\rm b}$: measured vs predicted \citep{Farago:2010} for all particles in our simulation.}
	\label{fig:P_P}
\end{figure}

We calculate the expected period (Eq~\ref{eq:T_farago}) for each test particle that we sample, and in Figure~\ref{fig:P_P} we plot a straight one-to-one comparison between the predicted and measured periods of precession.

Figure~\ref{fig:P_P} shows two populations --- the upper right group consists of test particles from our radius $50a_{\rm b}$ simulations, hence the larger periods, whereas the lower left group represents every test particle from the main body of our simulations (1.5$a_{\rm b} \leq a \leq 10a_{\rm b}$, summarised in Table~\ref{table:dynamics sims phase-space sampling}).

The scatter in period at the upper end of the groups of Figure~\ref{fig:P_P} is due to these periods being significantly greater than the duration of the simulation, and hence badly fitted. The scatter at lower periods ($P \lesssim 10^3 T_{\rm b} $) is due to deviations of the experimental values from the quadrupolar modal at low radii. But within the well behaved boxed region of Figure~\ref{fig:P_P}: 

$$ P_{\rm FL} / P_{\rm measured} = 1.008 \pm 0.029. $$

The results of our numerical investigation into the dynamics of circumbinary orbits in the elliptically restricted three body problem have correlated well with the analytic work of \citet{Farago:2010}. We have also explored the limits of their model due to the quadrupolar approximation. 

\section{Stability of circumbinary orbits} \label{sec:stability}

Figure~\ref{fig:log_period} shows test particle traces on the $(i\cos{W},i\sin{W})$ surface of section for our simulation of binary eccentricity $e_{\rm b}=0.5$ and binary mass fraction $\alpha_{\rm b} = 0.5$, as a function of test particle orbital radius $a/a_{\rm b}$ from the centre of mass of the binary. Each radius bin is equally sampled but we only plot orbits which remained stable throughout the simulation. It can be seen that at a radius of only $3a_{\rm b}$ very few of the initial test particles are actually stable, but at greater distances from the binary the surface of section fills in.

Figure~\ref{fig:log_period} shows that in the plotted simulation ($e_{\rm b}=0.5$, $\alpha_{\rm b} = 0.5$) the closest stable orbits to the binary are those at the centres of the islands of libration. These orbits are stable at a radius of $3a_{\rm b}$, with regions closer to the separatrix becoming stable out to $4a_{\rm b}$. As we move further away from the binary the retrograde orbits begin to acquire stability for radii $\gtrsim 5a_{\rm b}$, before the prograde orbits finally become stable at $\gtrsim 6a_{\rm b}$.

The time-averaged quadrupolar model of \citet{Farago:2010} predicts the dynamics of circumbinary orbits but can make no attempt to discern whether these orbits are viable. In the following sections we reveal characteristics which are due to resonances between the binary and test particle orbital periods which the \citeauthor{Farago:2010} model cannot explain due to the time-averaging carried out in its derivation.

\subsection{Suite of simulations}

We ran a suite of simulations to investigate the stability of circumbinary orbits. As discussed in \S~\ref{sec:separatrix and critical angle} the dynamics of the circumbinary phase space are such that every orbit crosses the $W=\pm\pi/2$ axis and there exists a symmetry which reflects $W=+\pi/2$ onto $W=-\pi/2$. We may therefore narrow down the region of phase space which we sample to only one value of the longitude of the ascending node, $W=+\pi/2$.

We run these simulations to 50,000 binary orbital periods, sampling phase space to a good density in orbital radius $a$ and inclination $i$, as laid out in Table~\ref{table:stability sims phase-space sampling}. We sample across binary eccentricity and mass fraction as in the above sections (see Table~\ref{table:binary-space sampling}).

\begin{table}
\caption{Sampling of circumbinary phase space where \newline $a_{\rm b}$ = binary semi-major axis and $T_{\rm b}$ = binary orbital period. }
\label{table:stability sims phase-space sampling}
\begin{center} 
\begin{tabular}{c c c c}
Orbital Element & min & max & $\triangle$ \\ 
\hline
semi-major axis $a$	& $\leq1.5a_{\rm b}$ & $\geq5a_{\rm b}$ & 0.05$a_{\rm b}$\\
inclination $i$ & $0$ & $\pi$ & $\pi/80$ \\
longt. of the asc. node $W$ & $\pi/2$ & $\pi/2$ & $-$ \\
true anomaly $v$ & $0$ & $2\pi$ & $\pi/3$ \\
\hline
sim. length and snapshot $\triangle t$ &  & $5\times10^4T_{\rm b}$ & $200T_{\rm b}$ 
\end{tabular}
\end{center} 
\end{table}

\subsection{A measure of stability} \label{sec:describe stability}

As discussed in \S~\ref{sec:numerial setup}, each test particle is monitored for instability. Unstable orbits are identified and removed during integration where a test particle is perturbed sufficiently from its initial orbit to approach either star, or if it evolves onto an unbound trajectory ($e>1$). Post-simulation stability criteria are applied to identify and reject test particles which do not quite reach escape velocity.

\begin{figure}
	\centering
	\includegraphics[width=\columnwidth]{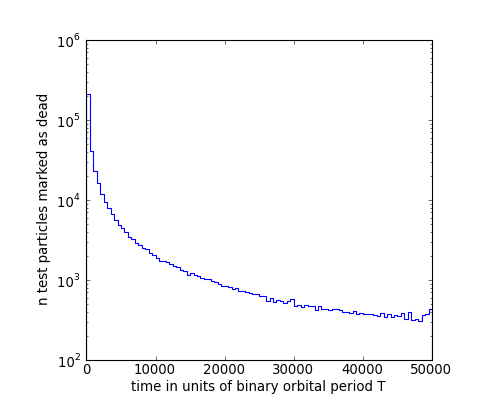}
	\caption{A histogram of the escape time $t_{\rm escape}$ for particles which sample unstable orbits.}
	\label{fig:hist t_death}
\end{figure}

Whereas in section~\ref{sec:dynamics} we were concerned with the orbits which survived the simulation, here we are more interested in those which don't. In Figure~\ref{fig:hist t_death} we plot a histogram of the escape times $t_{\rm escape}$ at which particles are rejected from the simulation. The vast majority of unstable particles are caught at the start of the simulation --- of the approximately half a million unstable test particles, over half of these are caught within the first 1000 binary orbital periods, with the distribution tailing off steeply even in log-space.

Since we sample each orbit from multiple initial values of the true anomaly $v$ (Table~\ref{table:binary-space sampling}) we measure an orbit's long-term stability by the fraction of initial test particles which survive the simulated duration of $5\times10^4T_{\rm b}$.

\begin{figure*}
	\centering
	\includegraphics[height=0.92\textheight]{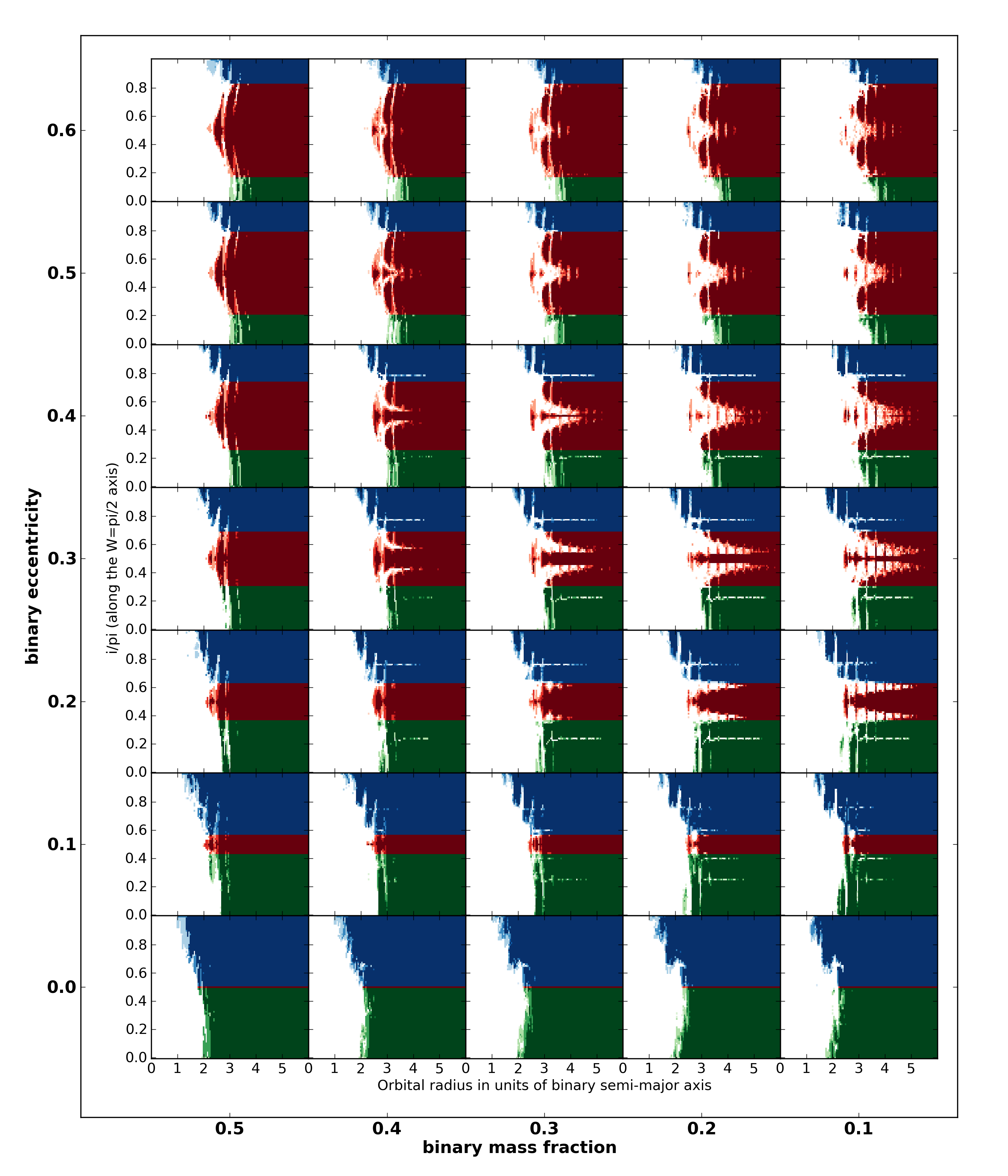}
	\caption{Orbital stability plotted as a function radius $a/a_{\rm b}$ and inclination $i$ on the $W=\pi/2$ axis, across binary eccentricity --- mass fraction parameter space. Colours: \newline
	Green: prograde ($i<\pi/2$) \newline 
	Blue: retrograde ($i>\pi/2$) \newline
	Red: island of libration centred at ($i=\pi/2,W=\pi/2$) 
	}
	\label{fig:stability}
\end{figure*}

In Figure~\ref{fig:stability} we show a density plot of this measure of stability across our entire parameter space. The major axes of this figure correspond to the simulation parameters of binary eccentricity and mass fraction, whilst the minor axes correspond to orbital radius and inclination. This is a density plot where each pixel corresponds to a sampled orbit, and the transparency to our measure of stability --- a darker colour indicates a more stable orbit. We preserve the colour scheme of Figure \ref{fig:surface of section}.

We draw the reader's attention to the following features of Figure~\ref{fig:stability}, as revealed by our exquisitely detailed simulations:

\begin{enumerate}
\item First, and very broadly speaking, orbits are more stable at lower binary eccentricity $e_{\rm b}$. 
\item With equal generality, retrograde orbits (blue) appear to be the most stable, followed by librating orbits (red), and finally prograde orbits (green). The difference in radius of the innermost stable orbit across inclination can be as large as $2a_{\rm b}$.
\item There are vertical striations of instability, most noticeable in the higher eccentricity simulations $e_{\rm b} \geq 0.5$, and predominantly at inclinations of $i \sim 0$ and $i \sim \pi/2$. We hypothesize that these regions of instability are due to orbital resonances between a test particle and the binary.
\item We note very thin horizontal {\it pinnacles} of instability in non-librating orbits $0.2 \leq e_{\rm b} \leq 0.4$. These pinnacles are located at inclinations $i\approx\pi/4$ and $i\approx 3\pi/4$, and extend up to $3a_{\rm b}$ into otherwise stable phase space.
\item More central to the pinnacles discussed above are wider {\it peninsulas} of instability in the librating region. These appear symmetrically either side of $i=\pi/2$ in the librating region for simulations of $e_{\rm b} \geq 0.2$ and converge upon each other as $e_{\rm b} \rightarrow 0.6$.
\item The horizontal pinnacles and peninsulas are a function of binary mass fraction $\alpha_{\rm b}$. These features do not appear in the $\alpha_{\rm b}=0.5$ simulations, and are magnified towards increasingly extreme values of $\alpha_{\rm b}$.
\end{enumerate}

\subsection{Previous work}

There exist numerous papers in the literature in which authors investigate long-term orbital stability within the coplanar circular restricted three body problem, both numerically and analytically, and with emphasis on circumstellar and circumbinary orbits.

With improvements in computation power authors have relaxed the circular constraint in the problem to investigate eccentric binary systems \citep{Dvorak:1989,Holman:1999,Musielak:2005}. But only recently have we had the computation power to relax the coplanar constraint on the problem to investigate inclined orbits. 

\citet{Pilat-Lohinger:2003} performed three dimensional numerical experiments to determine inclined stability but they did not explore the complex libration structure of the phase space\footnote{The inclined simulations of \citeauthor{Pilat-Lohinger:2003} contain only test particles of initial longitude of the ascending node $W=0$ (\citeauthor{Pilat-Lohinger:2003}, private communication). As such they did not sample any librating orbits, which do not intersect the $W=0$ or $W=\pi$ axes.}. \citeauthor{Pilat-Lohinger:2003} also considered only prograde inclinations up to $50^{\circ}$ and systems of equal mass fraction ($\alpha_{\rm b}=0.5$).

This paper presents the first dynamic-aware analysis of the stability of inclined circumbinary orbits throughout binary mass fraction --- eccentricity parameter space.


\subsection{Escape time}

\begin{figure*}
	\centering
	\includegraphics[height=0.92\textheight]{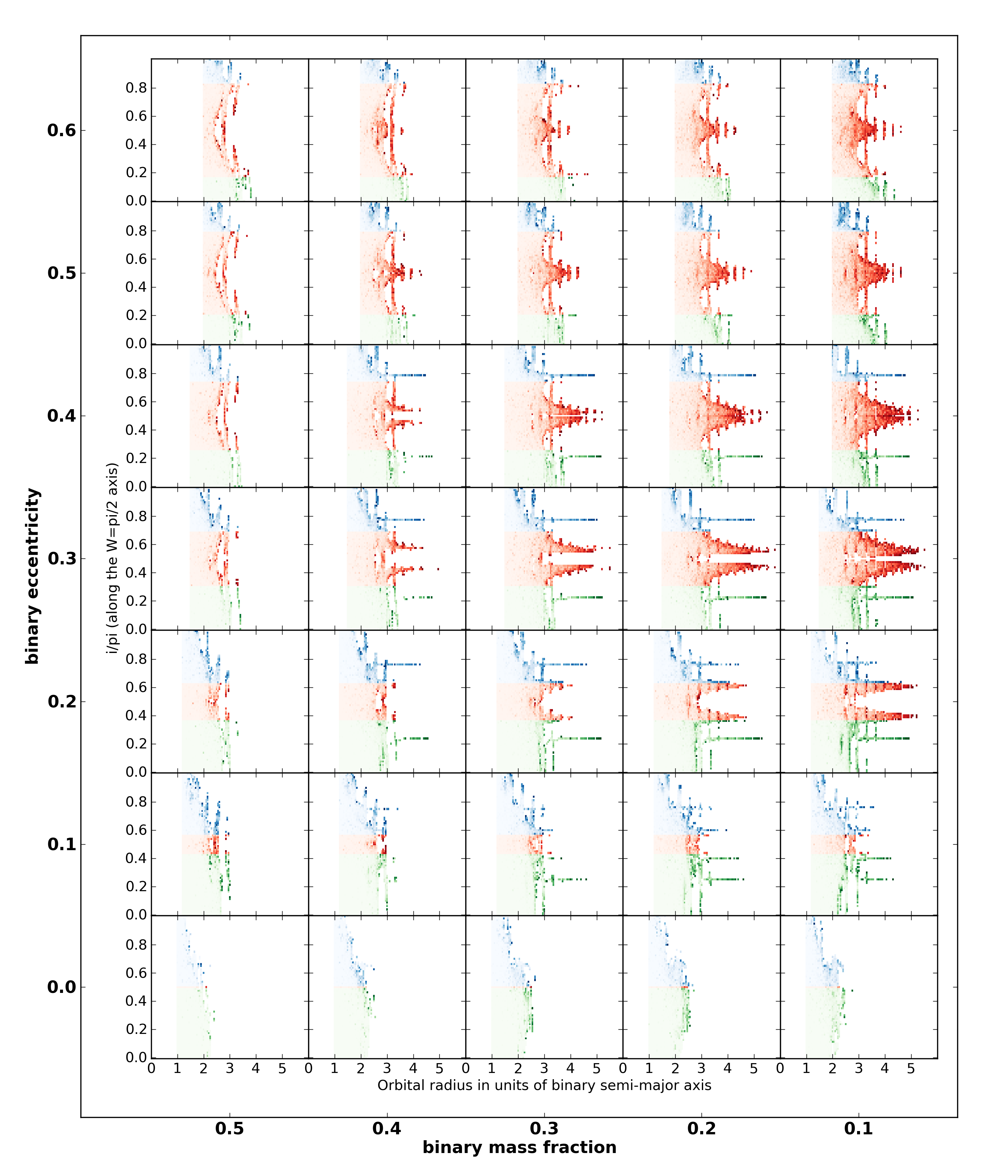}
	\caption{Inverse escape time $1/t_{\rm escape}$ as a function of radius $a/a_{\rm b}$ and inclination $i$ on the $W=\pi/2$ axis, across binary eccentricity --- mass fraction parameter space. Colours: \newline
	Green: prograde ($i<\pi/2$) \newline 
	Blue: retrograde ($i>\pi/2$) \newline
	Red: island of libration centred at ($i=\pi/2,W=\pi/2$) 
	}
	\label{fig:escape time}
\end{figure*}

In a companion figure to the stability plot of Figure~\ref{fig:stability} we show the escape time of each unstable orbit in Figure~\ref{fig:escape time}. Here we use the same axes as Figure~\ref{fig:stability} --- with major axes corresponding to the simulation parameters of binary eccentricity and mass fraction, and minor axes corresponding to orbital radius and inclination. This is a density plot where each pixel corresponds to an orbit sampled, and the transparency to the inverse of escape time $1/t_{\rm escape}$ --- a darker colour indicates a longer surviving orbit. We again preserve the colour scheme of Figure \ref{fig:surface of section}. Since we consider multiple test particles per sampled orbit we take an average for the escape time.

In Figure~\ref{fig:escape time} we find matching features to those in Figure~\ref{fig:stability}, as described above in \S~\ref{sec:describe stability}. But here we can also observe how long-lived the unstable orbits are. For example, in the low binary eccentricity simulations $e_{\rm b}<0.2$ we find that orbits are either very quickly unstable, or definitely stable. But for the simulations of higher binary eccentricity $e_{\rm b}$ and more imbalanced binary mass fraction $\alpha_{\rm b}$, the most striking features are long-lived. The test particles which sample phase space at these extreme points are almost stable, and survive for times of order the simulated duration $\sim 50,000T_{\rm b}$ (Table~\ref{table:stability sims phase-space sampling}).

Test particles on unstable orbits further inside the unstable region (closer to the binary) are quickly accreted onto the binary stars and/or ejected from the system. It is within this region of instability that inflows and outflows are likely to be important.

\subsection{Further considerations}

The phase space underlying our study of stability within the three-body problem is arguably significantly chaotic. We believe that the features which we report are real and significant to the consideration of matter around binaries, but we advise caution before cranking up the phase-space resolution. We should consider the effect of perturbations away from {\CHANGED this idealised model}, such as the volume of the bodies considered and the feedback effect on the internal binary of a third body of non-negligible mass.

\section{Conclusions}

Our simulations show that inclined circumbinary orbits in the elliptically-restricted three-body problem demonstrate three distinct families of behaviour: close-to-coplanar prograde ($i\sim0$) and retrograde ($i\sim\pi$) orbits precess in the longitude of the ascending node, whilst close-to-polar orbits ($i\sim\pi/2$ and $W\sim\pm\pi/2$) have their longitude of the ascending node and inclination coupled to precess about the centre of an island of libration.

We have extracted the critical angle $i_{\rm crit}(e_{\rm b})$ of the separatrix along the critical $W=\pi/2$ axis between these regions of behaviour as a function of binary eccentricity.

We have shown that the analytic time-averaged quadrupolar model of \citet{Farago:2010} provides an excellent description of the behaviours of circumbinary orbits at radii $\geq50a_{\rm b}$. We have also shown that their model becomes inaccurate to greater than 1\% at orbital radii $\leq 5a_{\rm b}$, and especially in cases of high binary eccentricity.

With the first 3D dynamic-aware analysis of the stability of circumbinary orbits we have discovered that these orbits are surprisingly stable throughout binary mass fraction --- eccentricity parameter space. Our detailed simulations have put numerical limits on this stability and revealed complex structure in orbital radius --- inclination space.

Our work shows that circumbinary phase-space is rich and dynamic, full of remarkable and stable orbits which do not behave simply. We should not presume any given binary system to lack a circumbinary component unless otherwise demonstrated. Such a component may be a source of obscuration, emission, inflow or outflow.

\section*{Acknowledgments}

SD thanks STFC for a studentship and KMB thanks the Royal Society for a University Research Fellowship. {\CHANGED We thank John Magorrian for useful discussions and our referee for his careful reading and useful comments towards this manuscript.} 

\bibliographystyle{mn2e}

\bibliography{$HOME/Papers/mybib1}

\end{document}